\documentclass[journal = jctcce]{achemso}
\usepackage{chemformula} 
\usepackage[T1]{fontenc} 

\graphicspath{{./}}
\newcommand{\beq}{\begin{equation}}
\newcommand{\eeq}{\end{equation}}
\newcommand{\bea}{\begin{eqnarray}}
\newcommand{\eea}{\end{eqnarray}}

\def\br{{\mathbf{r}}}

\def\bk{{\mathbf{k}}}

\def\bp{{\mathbf{p}}}
\def\bR{{\mathbf{R}}}
\def\bJ{{\mathbf{J}}}
\def\bL{{\mathbf{L}}}
\def\bA{{\mathbf{A}}}
\def\bC{{\mathbf{C}}}
\def\bD{{\mathbf{D}}}
\def\bE{{\mathbf{E}}}
\def\bH{{\mathbf{H}}}
\def\bS{{\mathbf{S}}}
\def\bPi{{\mathbf{\Pi}}}
\def\bsigma{{\mathbf{\sigma}}}
\def\bepsilon{{\mathbf{\varepsilon}}}
\def\bbH{{\bar{H}}}
\def\bbS{{\bar{S}}}
\author{Qiang Xu}
\affiliation{Materials Science $\&$ Engineering Program, Department of Chemistry, and Department of Physics $\&$ Astronomy,  University of California-Riverside, Riverside, California 92521, United States}

\author{Mauro Del Ben}
\affiliation{Applied Mathematics $\&$ Computational Research Division, Lawrence Berkeley National Laboratory, Berkeley, California 94720, United States}

\author{Mahmut Sait Okyay}
\affiliation{Materials Science $\&$ Engineering Program, Department of Chemistry, and Department of Physics $\&$ Astronomy,  University of California-Riverside, Riverside, California 92521, United States}

\author{Min Choi}
\affiliation{Materials Science $\&$ Engineering Program, Department of Chemistry, and Department of Physics $\&$ Astronomy,  University of California-Riverside, Riverside, California 92521, United States}

\author{Khaled Z. Ibrahim}
\affiliation{Applied Mathematics $\&$ Computational Research Division, Lawrence Berkeley National Laboratory, Berkeley, California 94720, United States}

\author{Bryan M. Wong}
\email{bryan.wong@ucr.edu, Website: http://www.bmwong-group.com}
\affiliation{Materials Science $\&$ Engineering Program, Department of Chemistry, and Department of Physics $\&$ Astronomy, University of California-Riverside, Riverside, California 92521, United States}
\title[An \textsf{achemso} demo]
  {Velocity-gauge Real-time Time-dependent Density Functional Tight-binding for Large-scale Condensed Matter Systems}

\abbreviations{IR,NMR,UV}
\keywords{American Chemical Society, \LaTeX}

\begin{document}

\begin{tocentry}
\begin{center} 
\resizebox{41mm}{43mm}{\includegraphics{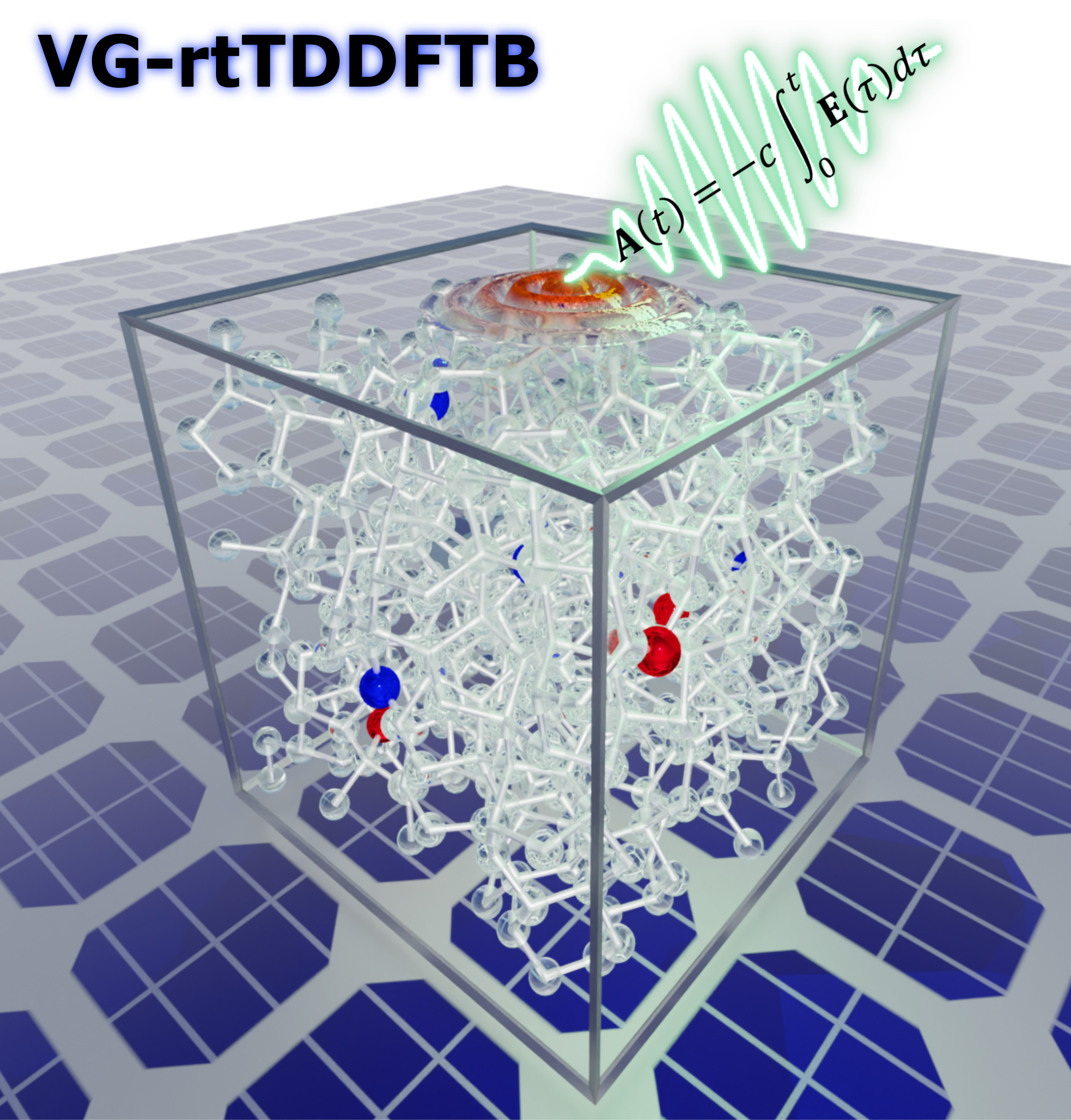}}
\end{center}
\end{tocentry}
\begin{abstract}
We present a new velocity-gauge real-time, time-dependent density functional tight-binding (VG-rtTDDFTB) implementation in the open-source DFTB+ software package (https://dftbplus.org) for probing electronic excitations in large, condensed matter systems. Our VG-rtTDDFTB approach enables real-time electron dynamics simulations of large, periodic, condensed matter systems containing thousands of atoms with a favorable computational scaling as a function of system size. We provide computational details and benchmark calculations to demonstrate its accuracy and computational parallelizability on a variety of large material systems. As a representative example, we calculate laser-induced electron dynamics in a 512-atom amorphous silicon supercell to highlight the large periodic systems that can be examined with our implementation. Taken together, our VG-rtTDDFTB approach enables new electron dynamics simulations of complex systems that require large periodic supercells, such as crystal defects, complex surfaces, nanowires, and amorphous materials. 


\end{abstract}

\section{Introduction}
Real-time time-dependent density functional theory (rtTDDFT)\cite{runge1984density,van1998causality,ullrich2011time} is a powerful approach for predicting the electron dynamics of quantum systems via solution of the time-dependent Kohn-Sham (KS) equations\cite{yabana1996time,yabana1999time,yabana1999application,bertsch2000real,meng2008real,provorse2016electron,goings2018real,pemmaraju2018velocity,meng2018real,li2020real}. rtTDDFT simulations can probe real-time electron dynamics in the presence of time-dependent external fields, which can be used to understand light-matter interactions and predict the linear/non-linear response of materials. The most common usage/implementation of rtTDDFT is for molecular (i.e., non-periodic) systems since the length gauge, which is used to calculate quantum dynamics in non-periodic systems, is relatively straightforward. However, the length-gauge formalism cannot be used for periodic systems since it breaks the translational symmetry of the Hamiltonian.\cite{yabana2006real,yabana2012time,mattiat2022comparison} For condensed matter systems, the time-dependent KS equations can be formally integrated under periodic boundary conditions\cite{bloch1929quantenmechanik} using the velocity-gauge (VG) formalism,\cite{pemmaraju2018velocity,mattiat2022comparison} which has been used to probe laser-induced electron dynamics in a variety of solid-state systems.\cite{andrade2015real,tancogne2020octopus,pemmaraju2018velocity,sato2014maxwell+,yabana2012time,krieger2015laser}


Despite its broad applicability, the immense computational expense of rtTDDFT calculations prohibits its use for 
large material systems such as crystal defects, complex surfaces, heterostructures, and amorphous systems, which require large supercells. To enable these large-scale electron dynamics simulations, an alternate theoretical formalism with a low computational cost is required. To address this need, we present the first velocity-gauge real-time time-dependent density functional tight-binding (VG-rtTDDFTB) implementation in the open-source DFTB+ software package (https://dftbplus.org) for large-scale and long-time electron dynamics simulations with periodic boundary conditions. Our VG-rtTDDFTB implementation makes use of the density functional tight-binding (DFTB) formalism\cite{elstner1998self,hourahine2020dftb+}, which is computationally efficient, relatively accurate, and scales extremely well with system size compared to full density functional theory (DFT).\cite{oviedo2016real,allec2019heterogeneous,plasmonic_dftb1,plasmonic_dftb2,convex_hull_dftb,molecules_dftb}

In this work, we derive the theoretical formalism and present a numerical implementation of VG-rtTDDFTB for electron dynamics simulations of large-scale condensed matter systems. Section 2 commences with a description of the VG-rtTDDFTB theoretical formalism. Section 3 provides computational details and benchmark calculations to demonstrate its accuracy and computational parallelizability on a variety of large material systems. Finally, we present an example of laser-induced electron dynamics in amorphous silicon to highlight the large periodic systems that can be examined with our approach, and  we conclude with future prospects/applications of our VG-rtTDDFTB implementation. 

\section{Theoretical formalism}
The time-dependent KS formalism with periodic boundary conditions in atomic units $(\hbar = e = m_{e} = 1)$ is given by
\beq
i\frac{\partial}{\partial t}|\psi_{n\bk}\rangle=\hat{H}_\text{KS}|\psi_{n\bk}\rangle,\label{eq:1}
\eeq

\noindent where $\psi_{n\bk}$ is the wavefunction in the Bloch representation. {In the long-wavelength limit, $\lambda{\gg}V^{1/3}$, where $\lambda$ and $V$ are the laser wavelength and volume of the simulation cell, respectively. Within this regime, the spatial variation of the electromagnetic field is negligible,\cite{ding2011gauge,cohen_louie_2016} which allows us to calculate macroscopic dielectric properties of solid-state systems\cite{baroni1986ab}. In the long-wavelength approximation, the VG-KS Hamiltonian is given by\cite{pemmaraju2018velocity,mattiat2022comparison,ding2011gauge}}
\bea
\hat{H}_\text{KS}&=&\frac{1}{2}\left[\hat{\textbf{p}}+\frac{1}{c}\bA (t)\right]^2+\hat{V}_\text{eff}[\rho]\nonumber \\
&=&\frac{1}{2}\hat{\bp}^2+\hat{V}_\text{eff}[\rho]+\underbrace{\frac{1}{c}\bA (t)\cdot\hat{\bp}+\frac{1}{2c^2}|\bA(t)|^2}_{\hat{H}_\text{ext}(t)},\label{eq:2}
\eea

\noindent where $\hat{\bp}$ and $\hat{V}_\text{eff}$ are the momentum and effective potential operators, respectively; $c$ denotes the speed of light in vacuum, $\rho(r)=\frac{1}{N_\bk}\sum_{n\bk}^\text{occ.}|\psi_{n\bk}(r)|^2$ is the electron density, ${N_\bk}$ is the number of $k$-points or the number of unit cells in the Born-von-Karmen supercell used in the calculation\cite{li2016large,xu2019ab}, and $\hat{H}_\text{ext}(t)$ is the Hamiltonian containing the external vector potential, which is given by
\beq
\bA(t)=-c\int_0^t\bE(\tau)d\tau,\label{eq:3}
\eeq
\noindent where $\bE$ is the external electric field; e.g., $\bE(t)=\bE_0\delta(t)$ for a delta function ``kick'' field, and $\bE(t)=\bE_0 \text{cos}(\omega t)$ corresponds to a monochromatic ``laser'' field with frequency $\omega$. {Note that the last term in Eq. (2) is a function of time that only gives rise to an overall phase factor of the form $e^{-i\int_0^t|A(\tau)|^2d\tau/(2c^2)}$ in the time-dependent wavefunction \cite{runge1984density}. This overall phase factor does not affect observables and density matrices in our implementation.}

The self-consistent-charge density functional tight-binding (SCC-DFTB) formalism\cite{elstner1998self} uses nonorthogonal pseudoatomic basis sets, $\{|\phi_\mu^{\zeta}\rangle\}$, and effective potentials in a two-center approximation. The collective index $\mu$ represents $( I, l, m)$ such that $|\phi_\mu^{\zeta}\rangle$ denotes the orbital centered on the $I$th atom of the $\zeta$th periodic cell image in real space with $(l,m)$ angular momentum quantum numbers, where $\phi_{\mu}(\br-\bR_I-\bL_\zeta)=\langle \br|\phi_{\mu}^{\zeta}\rangle$, $\bR_I$ and  $\bL_\zeta$ are the positions of the $I$th atom and $\zeta$th periodic image, respectively. The Bloch state in the SCC-DFTB formalism can be rewritten as
\beq
|\psi_{n\bk}\rangle=\sum_{\zeta}\sum_{\mu=1}^{N_b}C_{n\bk}^{\mu}e^{i\bk\cdot\bL_{\zeta}}|\phi_\mu^\zeta\rangle,\label{eq:4}
\eeq
\noindent where $N_b$ is the number of atomic basis functions in the unit cell. Combining Eqs.~(\ref{eq:1}) and (\ref{eq:4}), the time-dependent DFTB equations in matrix form becomes 
\beq
i\frac{\partial}{\partial t}\bC_{n\bk}=\bS_\bk^{-1}\bH_\bk\bC_{n\bk},\label{eq:5}
\eeq
\noindent where $\bC_{n\bk}=[C_{n\bk}^1,C_{n\bk}^2,...]^T$ is the coefficient vector of the $n\bk$th state. Alternatively, the equation of motion for the $k$-dependent density matrices can be obtained from Eq.(\ref{eq:5}) as
\beq
i\frac{\partial}{\partial t}\bD_{\bk}=\bS_\bk^{-1}\bH_\bk\bD_{\bk}-\bD_\bk\bH_\bk\bS_\bk^{-1},\label{eq:6}
\eeq

\noindent where the $k$-dependent density matrix is defined as $\bD_\bk=\sum_{n}^\text{occ.}\bC_{n\bk}\bC_{n\bk}^{\dagger}$, and the elements of the $k$-dependent overlap ($\bS_\bk$) and Hamiltonian ($\bH_\bk$) matrices are given by
\beq
S_\bk^{\mu\nu}=\sum_{\zeta} e^{-i\bk\cdot\bL_\zeta}\langle\phi_{\mu}^\zeta|\phi_\nu^{0}\rangle=\sum_{\zeta} e^{-i\bk\cdot\bL_\zeta}\bbS^{\mu\nu}(\bL_\zeta),\label{eq:7}
\eeq

\beq
H_\bk^{\mu\nu}=\sum_{\zeta} e^{-i\bk\cdot\bL_\zeta}\langle\phi_{\mu}^\zeta|\hat{H}_\text{DFTB}|\phi_\nu^{0}\rangle=\sum_\zeta e^{-i\bk\cdot\bL_\zeta}\bbH^{\mu\nu}(\bL_\zeta).\label{eq:8}
\eeq

To numerically evaluate the VG Hamiltonian, the $\bbH^{\mu\nu}(\bL_\zeta)$ term in Eq.~(\ref{eq:8}) can be rewritten within the SCC-DFTB approximation as

\beq
\bbH^{\mu\nu}(\bL_\zeta)=\bbH^{\mu\nu}_{(0)}(\bL_\zeta)+\bbH_{(2)}^{\mu\nu}(\bL_\zeta)+\bbH_\text{ext}^{\mu\nu}(\bL_\zeta),\label{eq:9}
\eeq

\noindent where $\bbH^{\mu\nu}_{(0)}(\bL_\zeta)\equiv\langle\phi_\mu^{\zeta}|\frac{1}{2}\hat{\bp}^2+\hat{V}_\text{eff}[\rho_0]|\phi_\nu^0\rangle$, $\bbH^{\mu\nu}_{(2)}(\bL_\zeta)$, and $\bbH^{\mu\nu}_\text{ext}(\bL_\zeta)$ include contributions from the non-self-consistent-charge, self-consistent-charge (SCC), and external potential terms in DFTB, respectively. In SCC-DFTB, all of the $\bbS^{\mu\nu}(\bL_\zeta)$ and $\bbH^{\mu\nu}_{(0)}(\bL_\zeta)$ matrix elements are pre-tabulated using Slater-Koster techniques\cite{slater1954simplified} with a reference density $\rho_0$. The diagonal elements of $\bbH^{\mu\nu}_{(0)}(\bL_\zeta)$  correspond to the atomic orbital energies, and the off-diagonal elements
are calculated in a two-centered approximation\cite{elstner1998self}. The SCC-term $\bbH^{\mu\nu}_{(2)}(\bL_\zeta)$ is defined as
\beq
\bbH_{(2)}^{\mu\nu}(\bL_\zeta)=\frac{1}{2}\bbS^{\mu\nu}(\bL_\zeta)\sum_{K}(\gamma_{IK}+\gamma_{JK})\Delta q_K,\label{eq:10}
\eeq

\noindent where Eq.~(\ref{eq:10}) depends on the Mulliken population $q_I=\frac{1}{2N_\bk}\sum_{n\bk}^\text{occ.}\sum_{\mu\in I,\nu}C_{n\bk}^{\mu*}S_\bk^{\mu\nu}C_{n\bk}^{\nu}+c.c.=\frac{1}{N_\bk}\sum_{\bk}\sum_{\mu\in I}[\bD_\bk\bS_\bk]^{\mu\mu}$, and $\Delta q_I= q_I-Z_I$ is the charge fluctuation with respect to the ion charge $Z_I$. The $\mu$th and $\nu$th basis indices denote orbitals centered on the $I$th and $J$th atoms, respectively.  $\gamma_{IJ}$ is defined as\cite{elstner1998self}
\beq
\gamma_{IJ}=\frac{1}{R_{IJ}}-S(U_I,U_J;R_{IJ}),\label{eq:11}
\eeq
\noindent where $R_{IJ}=|\bR_I-\bR_J|$ and $U_I$ is the chemical hardness or Hubbard parameter of the $I$th atom. The first term of Eq.~(\ref{eq:11}) is the long-range Coulomb interaction, and $S$ is the short-range term that decays exponentially\cite{elstner1998self}. For periodic boundary conditions, the long-range part of Eq.~(\ref{eq:10}) can be evaluated using the standard Ewald summation\cite{ewald1921evaluation,toukmaji1996ewald}, whereas the short-range part can be summed over a few neighbor periodic images of the central unit cell ($\zeta=0$). 

From Eq.~(\ref{eq:2}), the external potential term, $\bbH_\text{ext}^{\mu\nu}(\bL_\zeta)$, can be written as
\bea
\bbH_\text{ext}^{\mu\nu}(\bL_\zeta)&=&\langle\phi_\mu^\zeta|\hat{H}_\text{ext}|\phi_\nu^0\rangle\nonumber \\
&=&\frac{1}{c}\bA(t)\cdot\langle\phi_\mu^{\zeta}|\hat{\bp}|\phi_\nu^0\rangle+\frac{1}{2c^2}|\bA(t)|^2\bbS^{\mu\nu}(\bL_\zeta),\label{eq:12}
\eea

\noindent where the first term can be obtained from the expression $\langle\phi^\zeta_\mu|\hat{\bp}|\phi_\nu^0\rangle=-i\langle\phi^\zeta_\mu|\nabla\phi_\nu^0\rangle=i\langle\phi^\zeta_\mu|\nabla_J\phi_\nu^0\rangle$, $\nabla_J$ denotes the gradient with respect to  atomic position $\bR_J$, and $\nu \in J$th atom\cite{bonafe2020real}. The overlap matrix $\bS_\bk$ and Hamiltonian $\bH_\bk$ in Eqs.~(\ref{eq:5}) and (\ref{eq:6}) can be calculated using Eqs.~(\ref{eq:3},~\ref{eq:7}-\ref{eq:12}). The coefficient vectors or density matrices can be updated by 
integrating Eq. (\ref{eq:5}) or (\ref{eq:6}), respectively. In our implementation, we use the leapfrog algorithm to integrate Eq.~(\ref{eq:6}) for the $k$-dependent density matrix, $\bD_\bk$:
\beq
\bD_\bk(t+dt)=\bD_\bk(t-dt) + 2\dot\bD_\bk(t)dt.\label{eq:13}
\eeq
\noindent One can also integrate Eq.~(\ref{eq:5}) using the unitary Crank–Nicolson\cite{crank_nicolson_1947,castro2004propagators} time evolution of the coefficient vectors (which we have also implemented in our code). Our resulting VG-rtTDDFTB electron dynamics simulations give the time-dependent Mulliken charge and current $\bJ(t)$\cite{pemmaraju2018velocity}:

\beq
\bJ(t)=-\frac{1}{2\Omega N_\bk}\sum_{n\bk}^\text{occ.}\bC_{n\bk}^{\dagger}(t)\bPi_\bk(t)\bC_{n\bk}(t)+c.c.=-\frac{1}{\Omega N_\bk}\sum_\bk{\text{Tr}[\bD_\bk(t)\bPi_\bk(t)]},\label{eq:14}
\eeq
\noindent where $\Omega$ denotes the volume of the unit cell, and the elements of the current-momentum matrix is $\bPi_\bk^{\mu\nu}(t)=\sum_{\zeta}e^{-i\bk\cdot\bL_\zeta}\langle\phi_\mu^\zeta|\hat{\bp}|\phi_\nu^0\rangle+\frac{1}{c}\bA(t) S_\bk^{\mu\nu}$, which is a 3-component vector. The frequency-dependent conductivity, $\bsigma(\omega)$, and  dielectric function, $\bepsilon(\omega)$, can be derived using the time-dependent current generated by the $\delta$-function ``kick'' field in the linear-response regime\cite{pemmaraju2018velocity}:
\beq
\sigma_{ij}(\omega)=\frac{1}{E_{0j}}\int_0^T{e^{i\omega\tau}J_i(\tau)f(\tau)}d\tau,\label{eq:15}
\eeq
\beq
\varepsilon_{ij}(\omega)=1+\frac{4\pi i\sigma_{ij}(\omega)}{\omega}.\label{eq:16}
\eeq

\noindent where $f(\tau)=e^{-\tau/\tau_0}$ is a filtering function with  $\tau_0$ set to 200 a.u. in this work. With these quantities calculated, the absorption spectrum is given by the imaginary part of the dielectric function $\text{Im}[\varepsilon(\omega)]$. {Note that Eq. (15) does not contain any spatial dependence because of the long-wavelength approximation.\cite{cohen_louie_2016}}

\section{Results and Discussion}
\subsection{Computational details}
\begin{figure*}[!htb]
	\begin{center}		
		\includegraphics[width=1.0\textwidth]{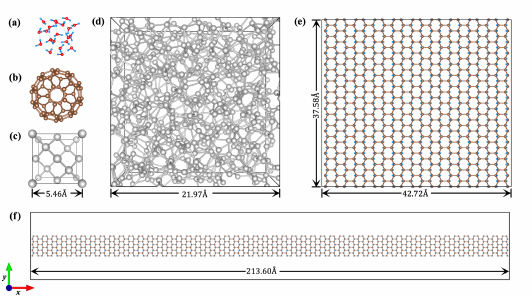}
		\caption{\label{fig:1} Molecular and periodic structures examined with our VG-rtTDDFTB approach: (a) (H${_2}$O)$_{21}$ cluster, (b) C$_{60}$, (c) c-Si, (d) a-Si, (e) 2D-graphane (C$_{600}$H$_{600}$), and (f) 1D-graphane (C$_{800}$H$_{800}$).}
	\end{center}
\end{figure*}

We first calculated the electronic ground state with SCC-DFTB using the DFTB+ package\cite{elstner1998self,hourahine2020dftb+}. The subsequent electron dynamics were then calculated with the VG-rtTDDFTB framework developed in this work. We performed electron dynamics simulations on representative chemical/material systems ranging from 0D (cluster) to 3D (bulk) geometries. Fig.~\ref{fig:1}(a) shows the optimized structure of a (H$_2$O)$_{21}$ cluster using the TIP5P potential\cite{mahoney2000five,james2005global,acuk}. The C$_{60}$ and crystal silicon (c-Si) structures shown in Figs.~\ref{fig:1} (b) and (c), respectively, were optimized with the SCC-DFTB method. The amorphous silicon (a-Si) structure containing 512 atoms in Fig.~\ref{fig:1}(d) was obtained from the last molecular dynamics step of GAP-MD simulation with a slow quench rate ($10^{11}$ K/s) from 1800 to 500 K\cite{deringer2018realistic}, which has been shown to generate reliable structures for a-Si compared to experiments\cite{deringer2018realistic}. For the partially periodic systems in Figs.~\ref{fig:1}(e) and (f), we built the 2D-graphane and 1D-graphane structures from the $10\times15\times1$ and $50\times4\times1$ supercells, respectively, of the boat conformer of graphane\cite{sofo2007graphane}. We used a $30$ \AA~vacuum space along the nonperiodic directions for all the clusters and partially periodic systems in this work.
 
For our DFTB/VG-rtTDDFTB calculations, we used the $mio$\cite{elstner1998self} and $pbc$\cite{sieck2003shape} Slater-Koster (SK) parameter sets for clustered and bulk/1D/2D systems, respectively. We found that the $siband$\cite{markov2015atomic,markov2015permittivity} SK set for c-Si and a-Si gave more accurate electronic structures and absorption spectra. We used maximum angular momenta of $s$, $p$, and $d$ for H, C/O, and Si elements, respectively. To benchmark our VG-rtTDDFT calculations, ground-state DFT and VG-rtTDDFT calculations were carried out with the real-space based OCTOPUS package\cite{andrade2015real,tancogne2020octopus} using the Perdew-Burke-Ernzerhof (PBE) exchange-correlation functional\cite{perdew1996generalized} with a 0.15 \AA~grid spacing for c-Si. For our absorption spectra calculations, we applied a $\delta$-pulse electric field of $E_x(t)=E_0\delta(t)$ and $E_0=0.005$ eV/\AA~applied along the $x$-axis at $t=0$ for all systems. The average current is subtracted from the total current induced by the kick field to reduce noise of the optical spectra. The 16$\times$ 16 $\times$ 16 $k$-point meshes were generated using the Monkhorst-Pack method\cite{monkhorst1976special} for c-Si, and a single $\Gamma$-point was used for the calculations of the other large-scale or cluster systems. We used a $0.002$ fs time step and the leapfrog integral method for all of our VG-rtTDDFTB electron dynamics simulations. Furthermore, we calculated the density of states (DOS) using a Gaussian function with a width of 0.05 eV for all cases. The Fermi level ($E_f$) for our DOS and band structure plots were shifted to a 0 eV reference energy. We defined our charge density using a Gaussian broadening of the atomic charges $\rho(\br;t)=\sum_I\frac{Q_I(t)}{\sqrt{(2\pi)^3}\eta^3}e^{-\frac{(\br-\bR_I)^2}{2\eta^2}}$, where $Q_I(t)=-\Delta q_I(t)$ denotes the $I$th atomic charge, and $\eta$ was set to $0.55$~\AA.

\subsection{Computational Results on Large Material Systems}

\begin{table*}	
	\caption{\label{tab:1}Walltime for various VG-rtTDDFTB simulations (18000 steps, 36 fs) on the NERSC \textit{Perlmutter} supercomputer.}
		\begin{tabular}{ccccccc}
		\hline	 & &  &  & number of  & number of & Walltime \\ 
    	System & Periodicity & SK set & $k$ mesh & basis functions & cores & (h) \\\hline
(H$_2$O)$_{21}$&0&$mio$&$1\times 1\times 1$&126&1&0.01\\

C$_{60}$&0&$mio$&$1\times 1\times 1$&240&1&0.06\\

c-Si&3&$siband$&$16\times 16\times 16$&72&64&0.35\\

a-Si&3&$siband$&$1\times 1\times 1$&4608&128&11.51\\
	
C$_{600}$H$_{600}$&2&$pbc$&$1\times 1\times 1$&3000&64&2.43\\

C$_{800}$H$_{800}$&1&$pbc$&$1\times 1\times 1$&4000&128&5.37
		\\ \hline
		\end{tabular}
\end{table*}

We implemented our VG-rtTDDFTB approach with a hybrid MPI/OpenMP parallelization to enable large-scale electron dynamics simulations. Our parallelization is accelerated by distributing the $k$-point index over MPI ranks because the electron dynamics simulation is largely independent of each $k$-point and requires only minimal inter-core communications. At the node level, for each $k$-point, the computational workloads are distributed among cores by using the multi-threaded OpenMP parallelization. To evaluate the computational efficiency for our parallelized VG-rtTDDFTB implementation, we simulated various systems for 18000 steps (36 fs) on the NERSC \textit{Perlmutter} supercomputer with a $\delta$-function ``kick'' field. Table~\ref{tab:1} shows that electron dynamics simulations of systems containing thousands of atoms can be efficiently performed on a modest computer cluster. These benchmarks show that our parallelized VG-rtTDDFTB implementation enables extremely efficient electron dynamics simulations of large complex condensed matter systems that are too computationally expensive with standard rtTDDFT approaches. 


\begin{figure*}[!htb]
	\begin{center}		
		\includegraphics[width=1.0\textwidth]{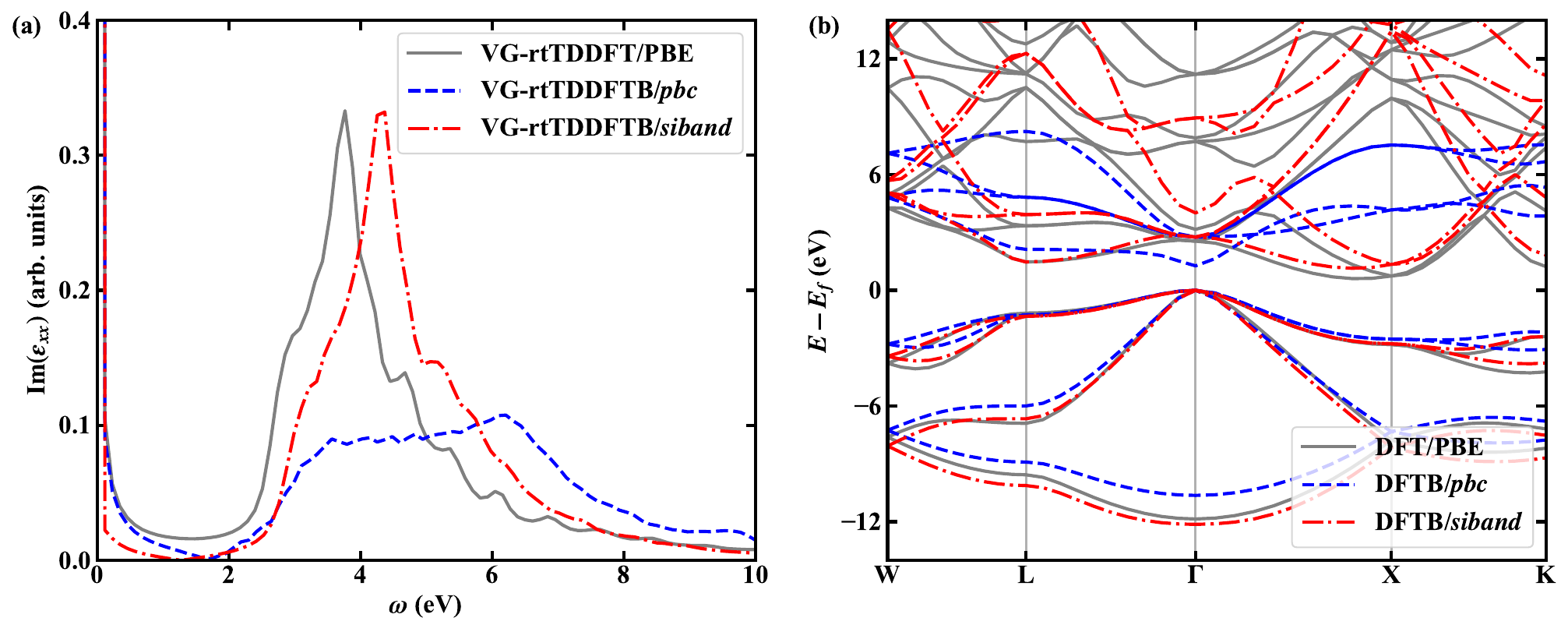}
		\caption{\label{fig:2} (a) Imaginary part of the dielectric function $\varepsilon_{xx}$ and (b) electronic band structures calculated with DFT (PBE) and DFTB using the $pbc$ and $siband$ Slater-Koster sets for crystalline Si.}
	\end{center}
\end{figure*}

To assess the accuracy of our VG-rtTDDFTB calculations for bulk systems, the optical spectra of c-Si were calculated by our VG-rtTDDFTB using $pbc$ and $siband$ SK sets in comparison with that by the VG-rtTDDFT using PBE functional. As shown in Fig.~\ref{fig:2} (a), the optical spectrum calculated by VG-rtTDDFTB using the $siband$ SK file are slightly blue-shifted, and generally consistent with full VG-rtTDDFT/PBE results. The VG-rtTDDFTB optical spectrum calculated with the $pbc$ parameter set, however, shows a significant discrepancy compared to the full VG-rtTDDFT/PBE results. The $siband$ parameter set was specifically constructed to accurately capture the electronic structure of bulk Si\cite{markov2015atomic}, which is necessary for reproducing its optical spectrum. As shown in Fig.~\ref{fig:2}(b), the DFTB band structures for bulk Si confirm the good agreement between the $siband$ parameterization and the DFT/PBE results. In contrast, the DFTB/$pbc$ band structure shows narrow conduction bands and significant discrepancies compared to the DFT/PBE result.

\begin{figure*}[!htb]
	\begin{center}		
		\includegraphics[width=1.0\textwidth]{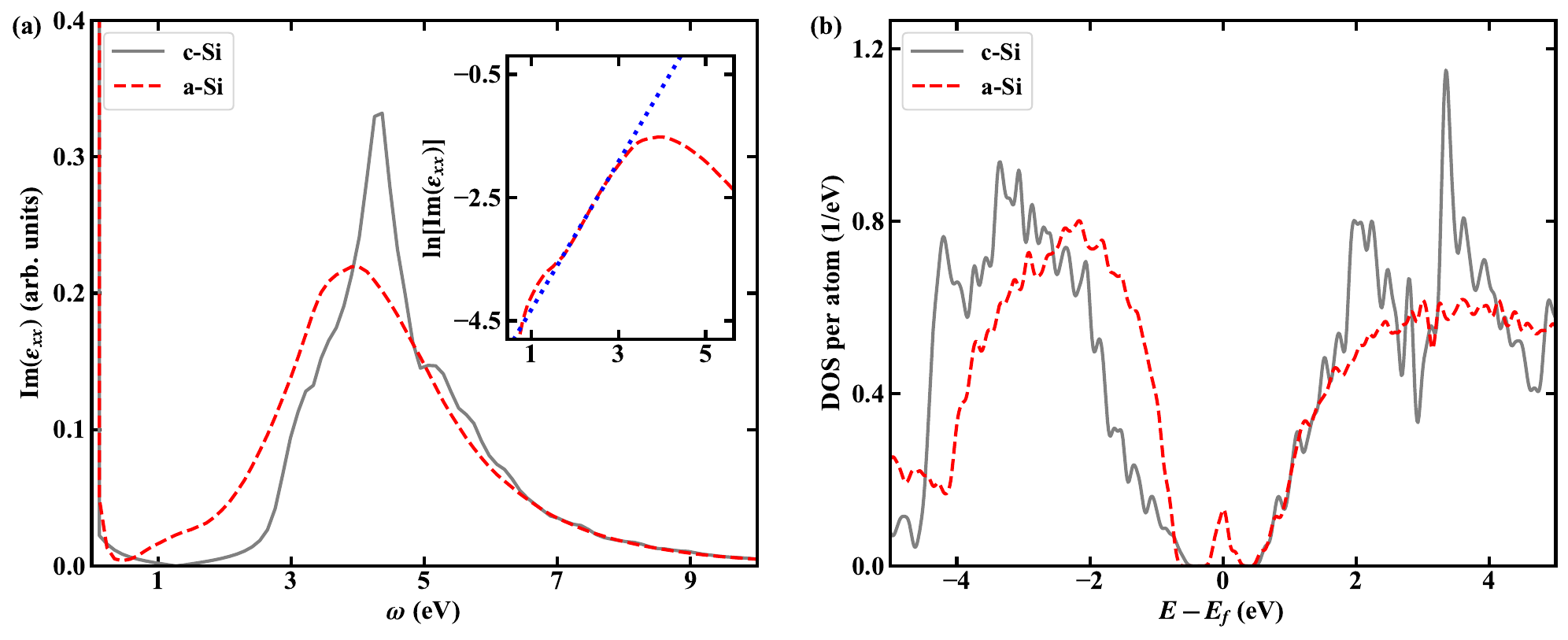}
		\caption{\label{fig:3} (a) Imaginary part of the dielectric function $\varepsilon_{xx}$ calculated by our VG-rtTDDFTB implementation. The dotted blue line in the inset is the least-squares fitting of the Urbach region using the expression ln$[\text{Im}(\varepsilon_{xx})]=\omega/{E_U}-5.52$ with $E_U=0.83$ eV. (b) density of states calculated with DFTB using the $siband$ Slater-Koster set for c-Si and a-Si.}
	\end{center}
\end{figure*}

Amorphous silicon has been widly investigated as a noncrystalline material with applications in solar cells\cite{carlson1976amorphous}, thin-film transistors\cite{powell1989physics}, and electrodes in batteries\cite{cui2009crystalline}. Despite its wide applicability, first-principles calculations for a-Si are rare because of the enormous computational expense of this amorphous system (which requires large supercells). To highlight the capabilities of our VG-rtTDDFTB approach, we calculate optical properties and electron dynamics for an a-Si structure shown in Fig~\ref{fig:1}(d). As shown in Fig.~\ref{fig:3}(a), the absorption edge of a-Si is broadened and slightly red-shifted compared to that of c-Si, which is generally consistent with the results reported in previous experiments\cite{grigorovici1968optical,pierce1972electronic,freeman1979optical}. Furthermore, we observed a clear ``Urbach'' absorption edge\cite{saito2000absorption} (from 1.7 to 2.8 eV) in the insert of Fig.~\ref{fig:3}(a), which arises from optical electron transitions between the localized (defected) and extended bands of the DOS shown in Fig.~\ref{fig:3}(b).

\begin{figure*}[!htb]
	\begin{center}		
		\includegraphics[width=1.0\textwidth]{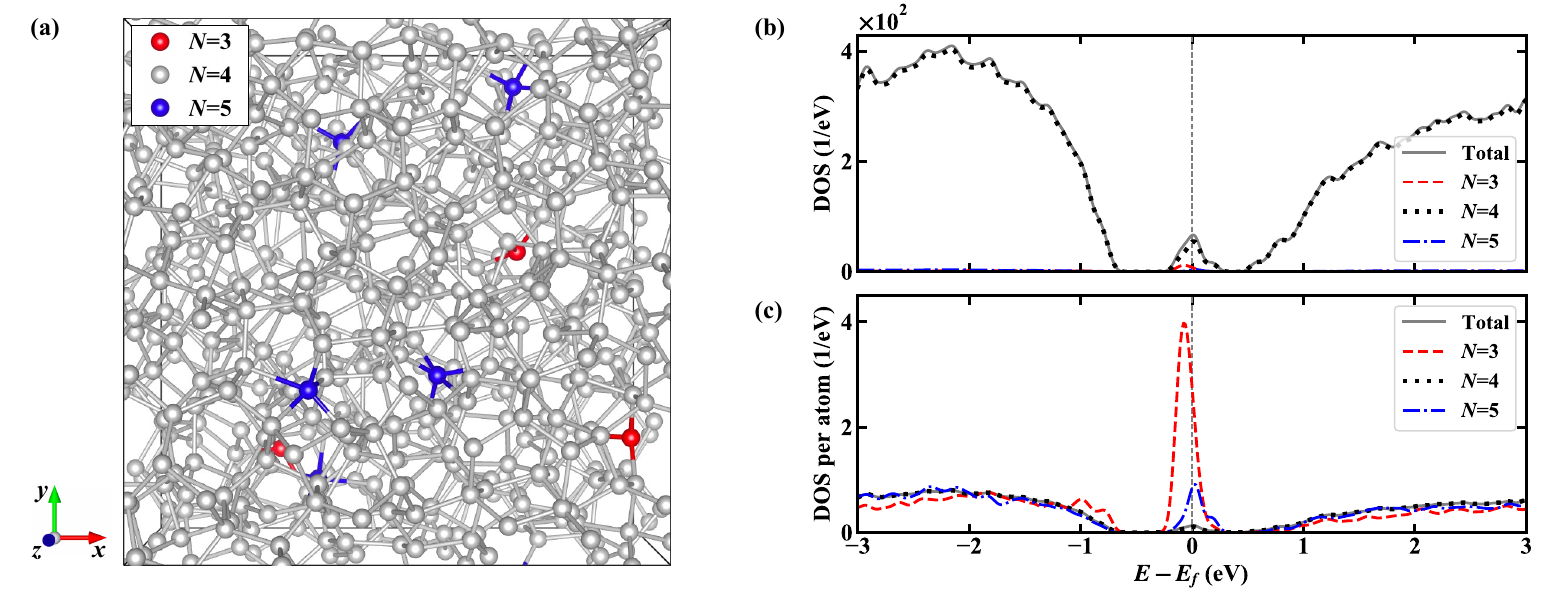}
		\caption{\label{fig:4} (a) Structure of amorphous Si containing 512 atoms with various $N$-coordinated atoms. Total/local (b) DOS and (c) DOS per atom for amorphous Si calculated using DFTB with $siband$ Slater-Koster set.}
	\end{center}
\end{figure*}
To investigate the effect of coordination defects on electron dynamics, we highlighted $N$-coordinated atoms for a-Si in Fig.~\ref{fig:4}(a) as the red (3-coordinated), silver (4-coordinated), and blue (5-coordinated) balls, which were determined by the bond-length cut-off of 2.85 \AA~ for each atom\cite{deringer2018realistic}.  Since 4-coordinated atoms are more prevalent (98.4\% of total atoms), the effect of other coordination defects in the total and local DOS [Fig.~\ref{fig:4}(b)] are less pronounced. However, the DOS per atom in Fig.~\ref{fig:4}(c) shows that coordination defects significantly influence the electronic structure near the Fermi level of a-Si. The 3-coordinated atoms have a large peak slightly below the Fermi level due to their localized unbonded character (i.e., dangling bonded atoms), whereas 5-coordinated atoms have a small peak above the Fermi level due to extra bonding, known as ``floating bonds.''\cite{pantelides1986PRL, bernstein2019Angew}. These states near the Fermi level have a significant influence on low-energy excitations and dynamics, which result in a broadened absorption spectrum.

\begin{figure*}[!htb]
	\begin{center}		
		\includegraphics[width=1.0\textwidth]{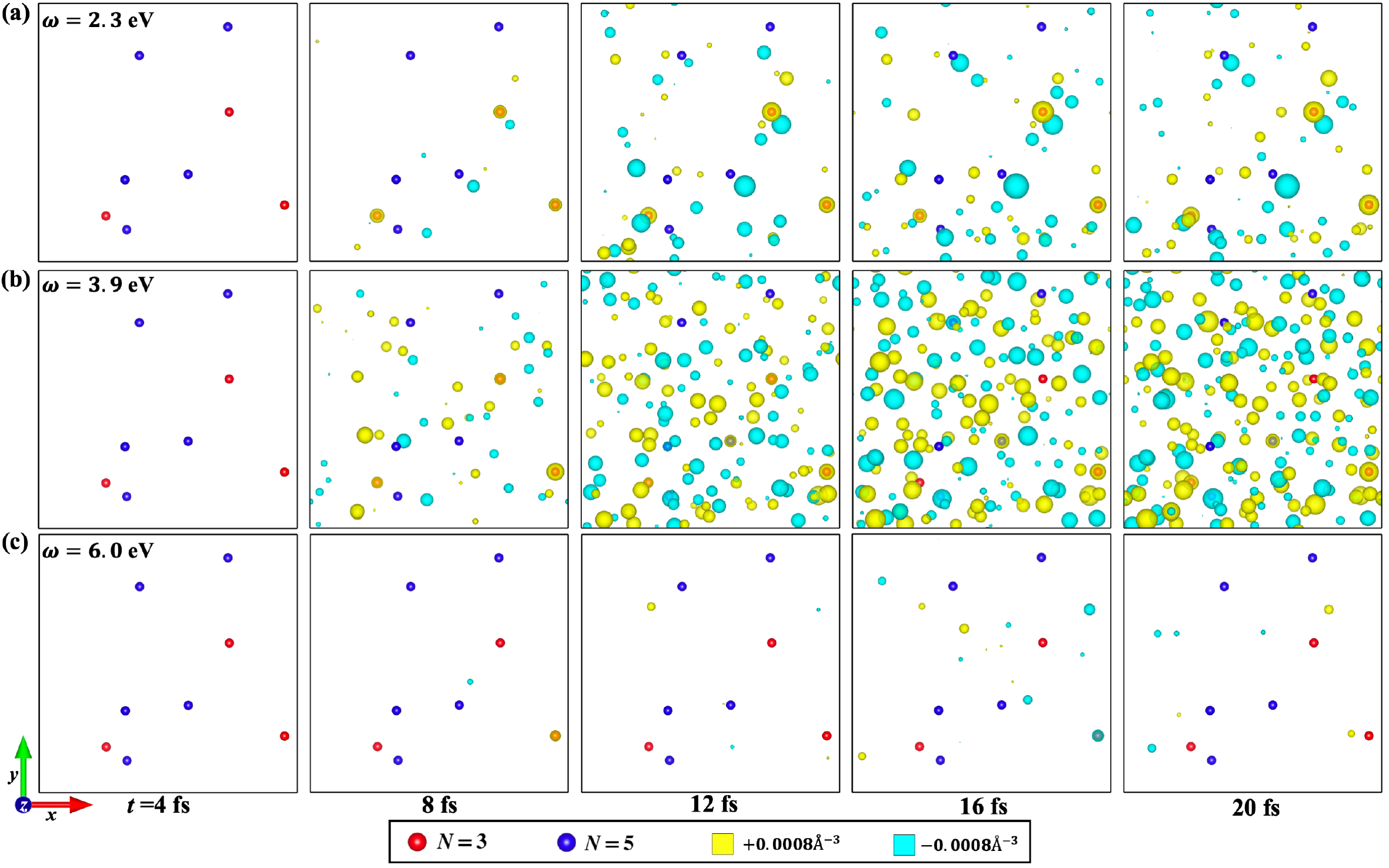}
		\caption{\label{fig:5} Snapshots of laser-induced total charge density fluctuations $\Delta\rho(\br;t)=\rho(\br;t)-\rho(\br;0)$ for $N$-coordinated atoms with respect to the ground state at laser frequencies of (a) 2.3, (b) 3.9, and (c) 6.0 eV. The 4-coordinated atoms are not shown for clarity.}
	\end{center}
\end{figure*}

\begin{figure*}[!htb]
	\begin{center}		
		\includegraphics[width=1.0\textwidth]{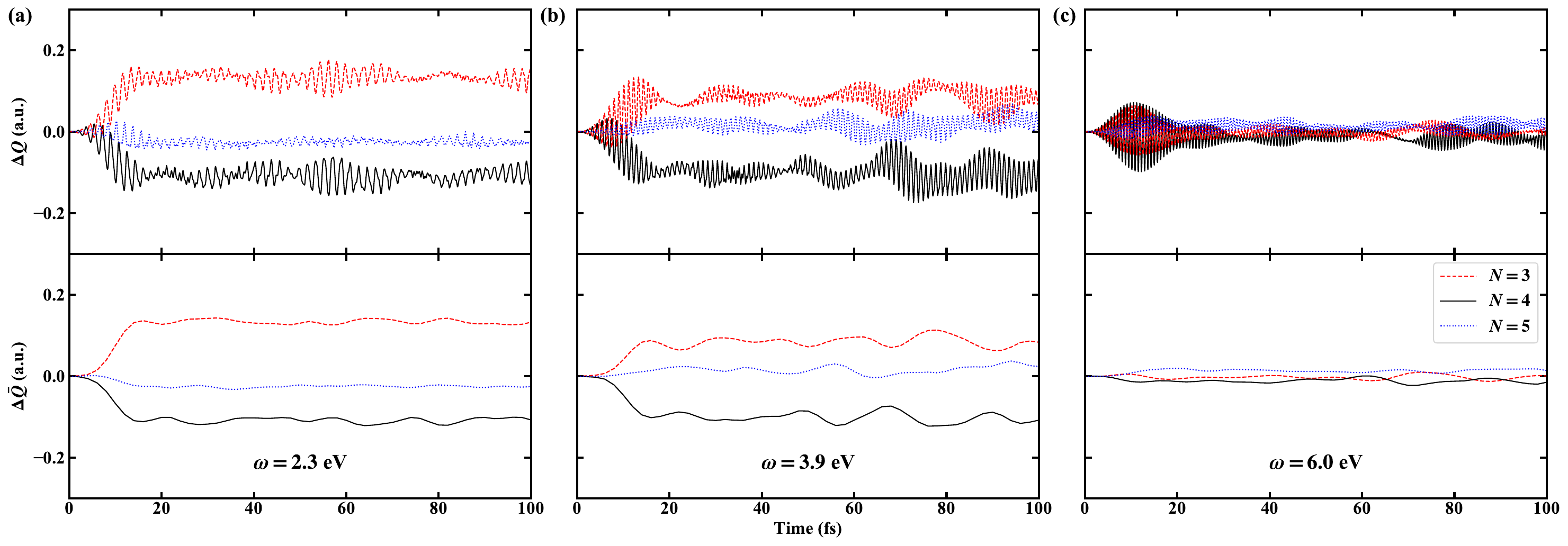}
		\caption{\label{fig:6} Laser-induced total charge evolution $\Delta Q(t)=Q(t)-Q(0)$ of various $N$-coordinated atoms at laser frequencies of (a) 2.3, (b) 3.9, and (c) 6.0 eV. The bottom panels depict the charge dynamics averaged every 2 femtoseconds.}
	\end{center}
\end{figure*}

To further study laser-induced electron dynamics of a-Si, we used a 20 fs sin$^2$-enveloped laser pulse (centered at 10 fs) with frequencies of 2.3, 3.9, and 6.0 eV, which corresponds to the edge, peak, and tail, respectively, of the absorption spectrum of a-Si as shown in Fig.~\ref{fig:3}(a). We applied an electric field along the $x$-axis with an intensity of $10^{12}$ W/cm$^2$ ($E_0\approx0.274$ eV/\AA) for our electron dynamics simulations. In Fig.~\ref{fig:5}, we present the instantaneous charge density with respect to the ground state of a-Si for different laser frequencies. We see that charge transfer under different frequencies are weak in early 4 fs due to the laser pulse is turning on the stage, whereas the charge transferred tends to maximize after 12 fs. Fig.~\ref{fig:5}(a) shows that low-frequency excitations at 2.3 eV induces electron transfer among the 3- and 4-coordinated atoms. Specifically, electrons are transferred from dangling-bonded ($N=3$) atoms for all snapshots after 8 fs. For medium-frequency excitations at 3.9 eV, Fig.~\ref{fig:5}(b) shows that more electron transfer occurs among the 4-coordinated atoms while electron transfer from 3-coordinated atoms to other atoms is still present. It is worth noting that  the different floating-bonded ($N=5$) atoms have different charge transfer modes at the same time. For high-frequency excitations at 6.0 eV, Fig.~\ref{fig:5}(c) shows that the least amount of charge transfer occurs. The 3-coordinated atoms lose electrons at 8 fs and subsequently gain charge at 16 fs. The overall electron dynamics are consistent with the absorption spectra peaks at each frequency shown in Fig.~\ref{fig:3}(a).

We calculated the Mulliken charge fluctuations among 3-, 4-, and 5-coordinated atoms with respect to the ground state  as shown in Fig.~\ref{fig:6}. For  low-frequency laser excitation (2.3 eV) shown in Fig.~\ref{fig:6}(a), the 3- and 5-coordinated atoms lose and gain electons, respectively, which is consistent with the electron transfer between localized (near Fermi level) and conduction bands in the Urbach region of the absorption edge\cite{saito2000absorption} as mentioned previously. For medium-frequency laser excitations (3.9 eV) shown in Fig.~\ref{fig:6}(b), the 5-coordinated atoms exhibit lose electrons, which is opposite of that shown in Fig.~\ref{fig:6}(a); in addition, electron transfer from 3-/4-coordinated atoms is weaker. These phenomena indicate that the extended valence band starts to transfer electrons to the extended conduction band. Fig.~\ref{fig:6}(c) shows that at high frequencies of 6.0 eV, the time-averaged electron transfer is relatively weak and comparable in magnitude among the $N$-coordinated atoms, which can be attributed to the high energy electron excitation between the extended valence and conduction bands that are far from Fermi level arising from the 4-coordinated atoms.

\section{Conclusion}
In summary, we have derived and implemented a new velocity gauge, real-time time-dependent density functional tight-binding (VG-rtTDDFTB) method in the open-source DFTB+ software package for electron dynamics simulations of large, condensed matter systems with periodic boundary conditions. Our VG-rtTDDFTB approach enables electron dynamics simulations of large condensed matter systems containing thousands of atoms with a favorable computational scaling as a function of system size. Our implementation uses a hybrid MPI/OpenMP parallelization scheme for massive parallelization to treat large systems on multi-core supercomputers. The computational efficiency of our VG-rtTDDFTB approach enables electron dynamics simulations of complex systems that require larger periodic supercells, such as crystal defects, complex surfaces, nanowires, and amorphous materials. 

In conclusion, we have demonstrated that our VG-rtTDDFTB implementation can execute large electron dynamics for periodic systems containing thousands of atoms. As a representative example, we performed a 100-fs electron dynamics simulation for amorphous silicon (containing 512 atoms) on a modest computer cluster to study laser-induced charge transfer dynamics. Our VG-rtTDDFTB calculations give mechanistic insight into time-resolved electron density fluctuations and electron transfer as the system is irradiated in real time with electromagnetic radiation. Our simulations also allow us to analyze different electron dynamics occurring between differently-coordinated atoms at various frequencies as the system is irradiated. We anticipate that our VG-rtTDDFTB approach could find broad usage for large periodic systems, particularly for material systems that are too large to handle with rtTDDFT. Further extensions to accelerate these calculations with specialized hardware accelerators \cite{fpga_dftb,allec2019heterogeneous,molecules_dftb} are currently underway in our group.

\begin{acknowledgement}
This work was supported by the U.S. Department of Energy, Office of Science, Office of Advanced Scientific Computing Research, Scientific Discovery through the Advanced Computing (SciDAC) program under Award Number DE-SC0022209. This research used resources of the National Energy Research Scientific Computing Center (NERSC), a U.S. Department of Energy Office of Science User Facility located at Lawrence Berkeley National Laboratory, operated under Contract No. DE-AC02-05CH11231 using NERSC award BES-ERCAP0023692. We gratefully acknowledge Dr. Nir Goldman for providing the ChIMES-modified \textit{siband} Slater-Koster file used to calculate VG-rtTDDFTB excited-state dynamics in this work. We gratefully acknowledge Dr. Steve D. Yang for constructing the table of contents figure used in this work.
\end{acknowledgement}





\nocite{1}
\bibliography{Refs}

\end{document}